\documentclass{iopjournal}

\usepackage{latexsym}
\usepackage{amssymb}
\usepackage{pdfpages}
\usepackage{graphics}
\usepackage{tikz}
\usepackage{tikz-feynman}

\begin{document}
\newcommand{\ja}{Jakuba\ss a-Amundsen }
\newcommand{\bfx}{\mbox{\boldmath $x$}}
\newcommand{\bfq}{\mbox{\boldmath $q$}}
\newcommand{\bfnabla}{\mbox{\boldmath $\nabla$}}
\newcommand{\bfalpha}{\mbox{\boldmath $\alpha$}}
\newcommand{\bfsigma}{\mbox{\boldmath $\sigma$}}
\newcommand{\bfeps}{\mbox{\boldmath $\epsilon$}}
\newcommand{\bfA}{\mbox{\boldmath $A$}}
\newcommand{\bfP}{\mbox{\boldmath $P$}}
\newcommand{\bfe}{\mbox{\boldmath $e$}}
\newcommand{\bfn}{\mbox{\boldmath $n$}}
\newcommand{\bfW}{{\mbox{\boldmath $W$}_{\!\!rad}}}
\newcommand{\bfM}{\mbox{\boldmath $M$}}
\newcommand{\bfI}{\mbox{\boldmath $I$}}
\newcommand{\bfJ}{\mbox{\boldmath $J$}}
\newcommand{\bfQ}{\mbox{\boldmath $Q$}}
\newcommand{\bfY}{\mbox{\boldmath $Y$}}
\newcommand{\bfp}{\mbox{\boldmath $p$}}
\newcommand{\bfk}{\mbox{\boldmath $k$}}
\newcommand{\bfks}{\mbox{{\scriptsize \boldmath $k$}}}
\newcommand{\bfqs}{\mbox{{\scriptsize \boldmath $q$}}}
\newcommand{\bfxs}{\mbox{{\scriptsize \boldmath $x$}}}
\newcommand{\bfalphas}{\mbox{{\scriptsize \boldmath $\alpha$}}}
\newcommand{\bfs}{\mbox{\boldmath $s$}_0}
\newcommand{\bfv}{\mbox{\boldmath $v$}}
\newcommand{\bfw}{\mbox{\boldmath $w$}}
\newcommand{\bfb}{\mbox{\boldmath $b$}}
\newcommand{\bfxi}{\mbox{\boldmath $\xi$}}
\newcommand{\bfzeta}{\mbox{\boldmath $\zeta$}}
\newcommand{\bfr}{\mbox{\boldmath $r$}}
\newcommand{\bfrs}{\mbox{{\scriptsize \boldmath $r$}}}
\newcommand{\bfps}{\mbox{{\scriptsize \boldmath $p$}}}

\renewcommand{\theequation}{\arabic{equation}}
\renewcommand{\thesection}{\arabic{section}}
\renewcommand{\thesubsection}{\arabic{section}.\arabic{subsection}}


\title{The parity-violating asymmetry including 
QED corrections in high-energy electron-nucleus collisions}

\author{Xavier Roca-Maza$^{1,2,3,4,*}$
  D.~H.~Jakubassa-Amundsen$^5$
  }

\affil{$^1$Departament de F\'isica Qu\`antica i Astrof\'isica, Mart\'i i Franqu\'es, 1, 08028 Barcelona, Spain}

\affil{$^2$Dipartimento di Fisica ``Aldo Pontremoli'', Universit\`a degli Studi di Milano, 20133 Milano, Italy}

\affil{$^3$INFN, Sezione di Milano, 20133 Milano, Italy}

\affil{$^4$Institut de Ci\`encies del Cosmos, Universitat de Barcelona, Mart\'i i Franqu\'es, 1, 08028 Barcelona, Spain}

\affil{$^5$Mathematics Institute, University of Munich, Theresienstrasse 39, 80333 Munich, Germany}

\affil{$^*$Author to whom any correspondence should be addressed.}

\email{xavier.roca.maza@fqa.ub.edu}


\begin{abstract}
The parity-violating asymmetry, accounting for the vector and axial-vector vertex plus self-energy correction as well as for vacuum polarization, is calculated nonperturbatively
by solving the corresponding Dirac equation for the electronic scattering states.
Investigating the nuclei $^{27}$Al, $^{48}$Ca and $^{208}$Pb at collision energies in the GeV region and at  forward scattering angles
matching the experimental geometries, it is found that the combined QED effects change the parity-violating asymmetry by less than one percent.
The same is true for $^{12}$C and $^{208}$Pb at an energy of 150 MeV.
\end{abstract}


The investigation of the parity-violating asymmetry $A_{\rm pv}$, both experimentally \cite{An22,Ad21,Ab12,Ad22} and theoretically \cite{Ru82,Ho98,Ba12,Ho14}, 
helps to determine the weak charge distribution and the weak form factor \cite{Ad22}. In turn,
neutron densities and neuton skin thicknesses can be extracted \cite{Ad21,Ho01,Re21}.
To this aim it is mandatory for theory  to consider the effect of different corrections to $A_{\rm pv}$.
One of those are dispersive corrections, estimated with the help of the $\gamma-Z$ box diagram \cite{Si10,Go11}.
Further corrections are due to QED effects.
Part of those have been considered \cite{Er03} by means of modifying the weak charge $Q_{\rm w}$ with respect to its tree level value \cite{GH09} as in \cite{Ad21,Ad22}.
However, this method does not account for the dependence of the form factors on momentum transfer $q$.

The influence of vacuum polarization on $A_{\rm pv}$ with the help of the Uehling potential was considered in \cite{MS05} within the semiclassical Eikonal approximation,
 and it was conjectured \cite{MS05} and recently proven \cite{RH26} that its axial-vector contribution is negligible for the analysis of current experimental data.
The vertex plus self-energy (vs) correction was calculated in \cite{RJ25}, together with vacuum polarization.
Only the vector vs corrrection was included in that work. 
However, it was recently pointed out by Reed and Horowitz \cite{RH26b} that for the vertex correction also the axial-vector contribution is important.
Their conjecture of a severe reduction of the QED corrections was verified by considering both vs parts nonperturbatively in some low-energy geometries \cite{JR26}.

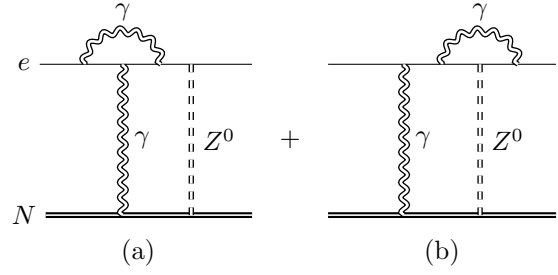
\begin{figure}[ht]
\centering
\begin{tikzpicture}

\begin{feynman}

\vertex (e1) at (-5, 1) {$e$};
\vertex (e2) at (-2, 1);
\vertex (n1) at (-5,-1) {$N$};
\vertex (n2) at (-2,-1);

\vertex (v1) at (-3.7, 1);
\vertex (v2) at (-3.7,-1);

\vertex (z1) at (-2.8, 1);
\vertex (z2) at (-2.8,-1);

\vertex (l1) at (-4.2,1);
\vertex (l2) at (-3.2,1);

\diagram*{

(e1) -- [plain] (e2),

(n1) -- [double, thick] (n2),

(v1) -- [photon, line width=2pt, edge label=$\gamma$] (v2),
(v1) -- [photon, line width=1pt, white, opacity=1] (v2),

(z1) -- [plain, dashed, line width=2pt, edge label=$Z^0$] (z2),

(z1) -- [plain, dashed, line width=1pt, white, opacity=1] (z2),

(l1) -- [photon, half left, looseness=1.5, line width=2pt, edge label=$\gamma$] (l2),
(l1) -- [photon, half left, looseness=1.5, , line width=1pt, white, opacity=1] (l2),
};

\end{feynman}
\node at (-1.5,0) {$+$};

\begin{feynman}

\vertex (e1r) at ( -1, 1);
\vertex (e2r) at ( 2, 1);
\vertex (n1r) at ( -1,-1);
\vertex (n2r) at ( 2,-1);

\vertex (v1r) at ( 0,-1);
\vertex (v2r) at ( 0, 1);

\vertex (z1r) at ( 1, 1);
\vertex (z2r) at ( 1,-1);

\vertex (p1) at ( 0.5,1);
\vertex (p2) at ( 1.5,1);

\diagram*{

(e1r) -- [plain] (e2r),

(n1r) -- [double, thick] (n2r),

(v2r) -- [photon, line width=2pt, edge label=$\gamma$] (v1r),
(v2r) -- [photon, line width=1pt, white, opacity=1] (v1r),

(z1r) -- [plain, dashed, line width=2pt, edge label=$Z^0$] (z2r),

(z1r) -- [plain, dashed, line width=1pt, white, opacity=1] (z2r),

(p1) -- [photon, half left, looseness=1.5, line width=2pt, edge label=$\gamma$] (p2),
(p1) -- [photon, half left, looseness=1.5, , line width=1pt, white, opacity=1] (p2),
};

\node at (-3.5, -1.5) {(a)};
\node at (0.5, -1.5) {(b)};
\end{feynman}

\end{tikzpicture}
\caption{Feynman diagrams for the vertex plus self-energy correction in the presence of photon and $Z^0$ exchange. (a) vector vs correction, (b) axial-vector vs correction. The double lines mark a non-perturbative interaction.\label{fig1}}
\end{figure}

In the present work, including again the vector vs and axial-vector vs parts, our nonperturbative
approach is applied to additional collision systems in order to compare the results with new ones from the semiclassical Eikonal theory \cite{RH26} and
to make further predictions.  The  Feynman diagrams for the vs correction are shown in Fig.~\ref{fig1}. The double boson lines in the diagrams indicate
that the corrections, considered in terms of the corresponding potentials, are included nonperturbatively as done for the vector vs correction in \cite{RJ25}.
 The vertex corrections show both ultraviolet (UV) and infrared (IR) divergences (see e.g. \cite{Va00} and references therein). The IR divergences are regularized using a small photon mass and, this term, cancels with the bremsstrahlung contribution. The UV divergence is regularized by adding the self-energy correction.

Like  the vector vs potential $V_{\rm vs}$, the axial-vector vs potential $A_{\rm vs}^{\rm ax}$ is constructed from the relation between the first-order
Born transition amplitude $A_{fi}$ and the underlying potential, which in case of the weak potential $A_{\rm w}$ \cite{Ho98} is given  (in atomic units, $\hbar =m_e =e =1$) by
$$
A_{fi}^{\rm w}\,=\,A_0 \int d\bfr \,e^{i\bfqs\bfrs}A_{\rm w}(r)$$
\begin{equation}\label{1}
=\;A_0\;\frac{G_Fc}{2 \sqrt{2}}\,\int d\bfr \,e^{i\bfqs\bfrs} \varrho_{\rm w}(r),
\end{equation}
where $A_0$ is the Born pre-factor (cf. Eq.(2.12) in Ref.~\cite{Jaku24}), $G_F$ the Fermi coup\-ling constant and $\varrho_{\rm w}$ the weak density.
We make use of the fact that the Born amplitude for the axial-vector vs correction is proportional to the uncorrected weak Born amplitude $A_{fi}^{\rm w}$,
\begin{equation}\label{2}
A_{fi}^{\rm w,vs}\,=\,F_1^{\rm w,vs} \;A_{fi}^{\rm w},
\end{equation}
where  $F_1^{\rm w,vs}$ is the weak vs form factor.  Its large-$q$ approximation, calculated by Reed and Horowitz \cite{RH26b}, reads 
\begin{equation}\label{3}
F_1^{\rm w,vs}(-q^2) =\frac{1}{2\pi c}\left\{ \frac12 \ln(-\frac{q^2}{c^2}) \,[3-\ln(-\frac{q^2}{c^2})] -1+\frac{\pi^2}{6}\right\}.
\end{equation}
From the inverse Fourier transform of (\ref{1}) by substituting $A_{fi}^{\rm w,vs}$ for $A_{fi}^{\rm w}$, one obtains the desired potential,
\begin{equation}\label{4}
A_{\rm vs}^{\rm ax}(r)\,=\,\frac{G_Fc}{4\sqrt{2} \pi^2} \int_0^\infty\!\!\!\! \bfq^2 \,d|\bfq| \,\frac{\sin(|\bfq|r)}{|\bfq|r}\,F_1^{\rm w,vs}(-q^2)\,F_{\rm w}(|\bfq|),
\end{equation}
where the weak form factor $F_{\rm w}$ is given by
\begin{equation}\label{5}
F_{\rm w}(q)\,=\,4\pi \int_0^\infty r^2dr\,j_0(qr)\,\varrho_{\rm w}(r),
\end{equation}
with $j_0$  a spherical Bessel function and $F_{\rm w}(0)=Q_{\rm w}$. 

In order to obtain $A_{\rm vs}^{\rm ax}(r)$ from Eq.~(\ref{4}), one needs $F_1^{\rm w,vs}$ at all $q$ values. In the large-$q$ limit, the expression in (\ref{3}) was employed. In order to check the relevance of low momentum transfers $(-q^2/c^2 <100) $, the expression for $F_1^{\rm vs}$ from the vector vs correction given in \cite{Jaku24} was used (being derived in \cite{Va00,BS19}), except for the constant -2 which was replaced by -1 
in order to guarantee continuity with (\ref{3}). However, this low-$q$ part in (\ref{4}) plays no role in the calculation of the $A_{\rm pv}$ at the kinematics relevant for the PREx and CREx experiments.

\begin{figure}
\vspace{-1.5cm}
\centering
\includegraphics[width=11cm]{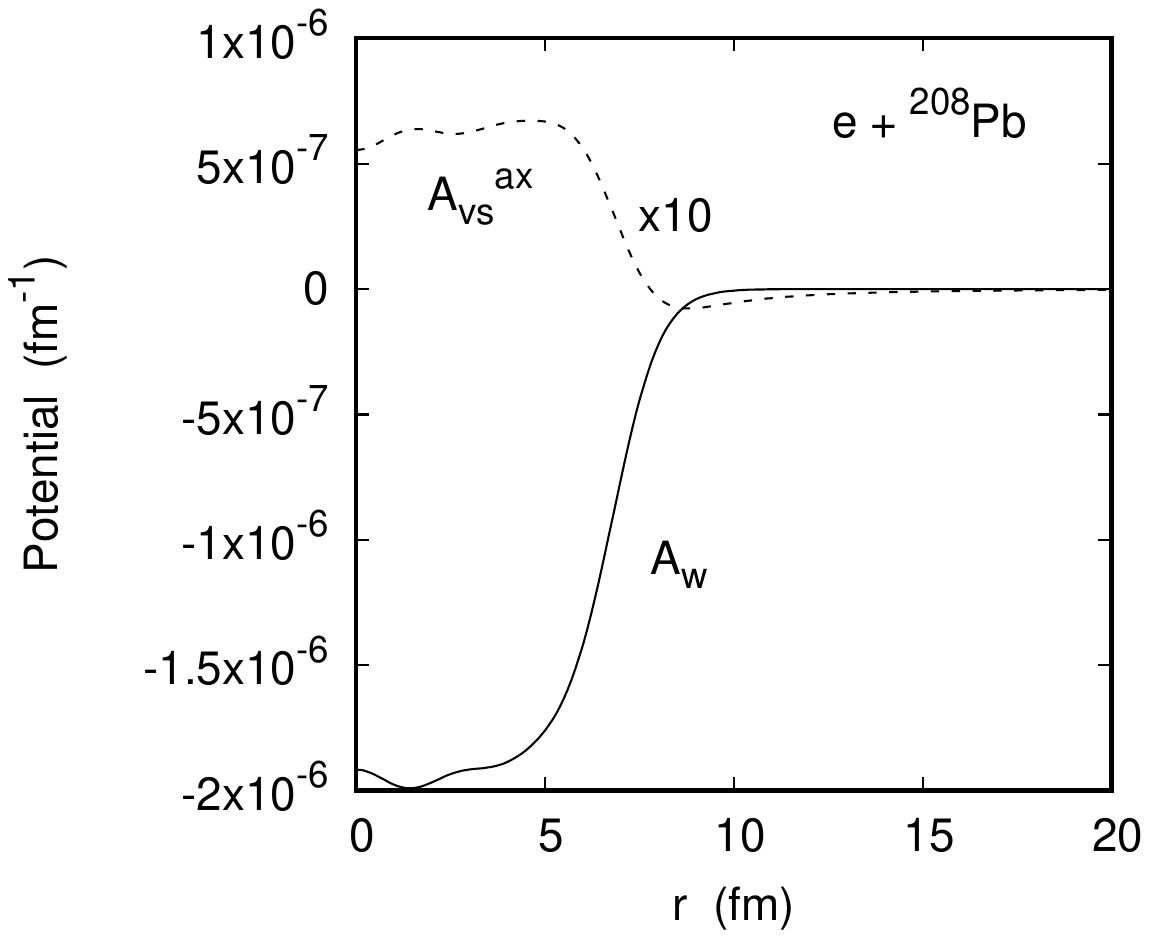}
\vspace{-0.5cm}
\caption
{
Spatial dependence of the weak potential $A_{\rm w}$ and the axial-vector vs potential $A_{\rm vs}^{\rm ax}$ (which is multiplied by a factor of 10) for $^{208}$Pb.
\label{fig2}}
\end{figure}

In Fig.~\ref{fig2}, $A_{\rm vs}^{\rm ax}(r)$ is compared to $A_{\rm w}(r)$.
The spatial dependence of the axial-vector vs potential is, except for the opposite sign, quite similar to the
behaviour of the weak potential, but it is nearly two orders of magnitude smaller.

Upon adding the axial-vector vs potential to the weak potential in terms of $\gamma_5(A_{\rm w} + A_{\rm vs}^{\rm ax}$), one arrives at the Dirac equation for the positive helicity ($\psi_+$), respectively negative helicity ($\psi_-$) scattering states,

\begin{equation}\label{6}
\left[ -ic\bfalpha\bfnabla + V_T(r) +V_{\rm vac}(r) +V_{\rm vs}(r)\right.
\left. \pm (A_{\rm w}(r) +A_{\rm vs}^{\rm ax}(r)\right)]\,\psi_\pm =E\,\psi_\pm,
\end{equation}
where $V_T$ is the target nuclear field, $V_{\rm vac}$ represents the Uehling potential as parametrized in \cite{FR76} and the potential $V_{\rm vs}$ describes the vector vs correction \cite{Jaku24}.
We note that the way of representing the QED effects in terms of potentials is approximate, but given the potentials the solution by means of the Dirac equation is exact,
comprising also diagrams like the one in Fig.10b of \cite{RH26}.

From the solutions of (\ref{6}), used in the subsequent partial-wave analysis, the QED-modified parity-violating asymmetry $A_{\rm pv}^{\rm QED}$ is obtained. Its relative change is defined by
\begin{equation}\label{7}
dA_{\rm pv}^{\rm QED}\,=\,A_{\rm pv}^{\rm QED} /A_{\rm pv} -1.
\end{equation}

\begin{figure}
\vspace{-0.5cm}
\centering
\includegraphics[width=11cm]{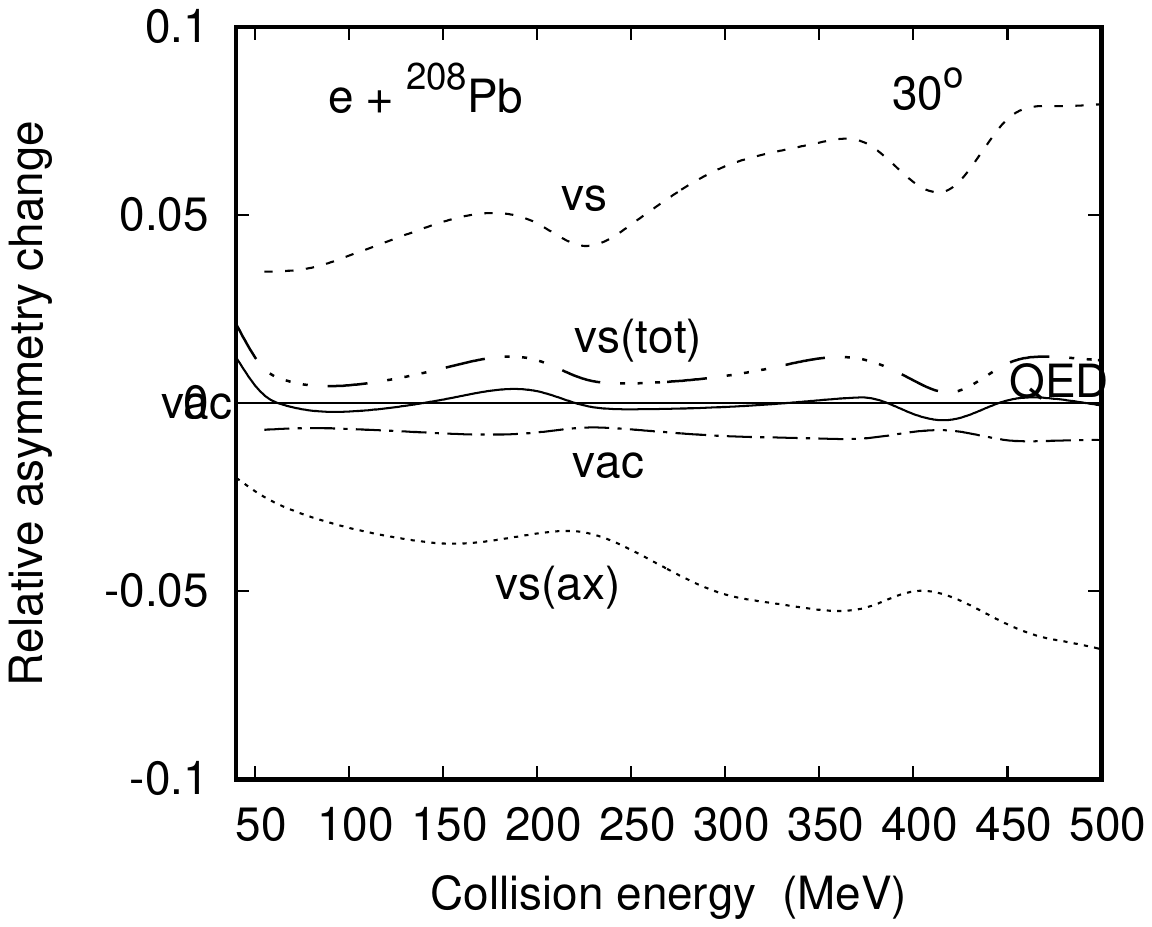}
\vspace{-0.5cm}
\caption
{
Relative spin asymmetry change $dA_{\rm pv}$ in $e+^{208}$Pb collisions at a scattering angle of $30^\circ$ as function of collision energy.
Shown are the QED changes by vacuum polarization $(-\cdot -\cdot -)$, by the vector vs correction $----$, by the axial-vector vs correction $(\cdots\cdots)$, by the combined vs correction $(-\cdots -)$, and  the total QED correction $dA_{\rm pv}^{\rm QED}$ (--------). \label{fig3}
}
\end{figure}

The energy dependence is displayed in Fig.~\ref{fig3} for electrons scattered from $^{208}$Pb at an angle of $30^\circ$.
Indeed it is found that $dA_{\rm pv}^{\rm QED}$ is at most  1\% at all energies considered.
Retaining only $V_{\rm vs}$ or $A_{\rm vs}^{\rm ax}$ in (\ref{6}) (in addition to $V_T$ and $A_{\rm w}$),
it is seen that the resulting $dA_{\rm pv}^{\rm ax}$ is nearly as large as $dA_{\rm pv}^{\rm vs}$ but of opposite sign, and that their
combined effect basically compensates the change from vacuum polarization at such forward angles.

\begin{figure}
\centering
\vspace{-1.5cm}
\includegraphics[width=11cm]{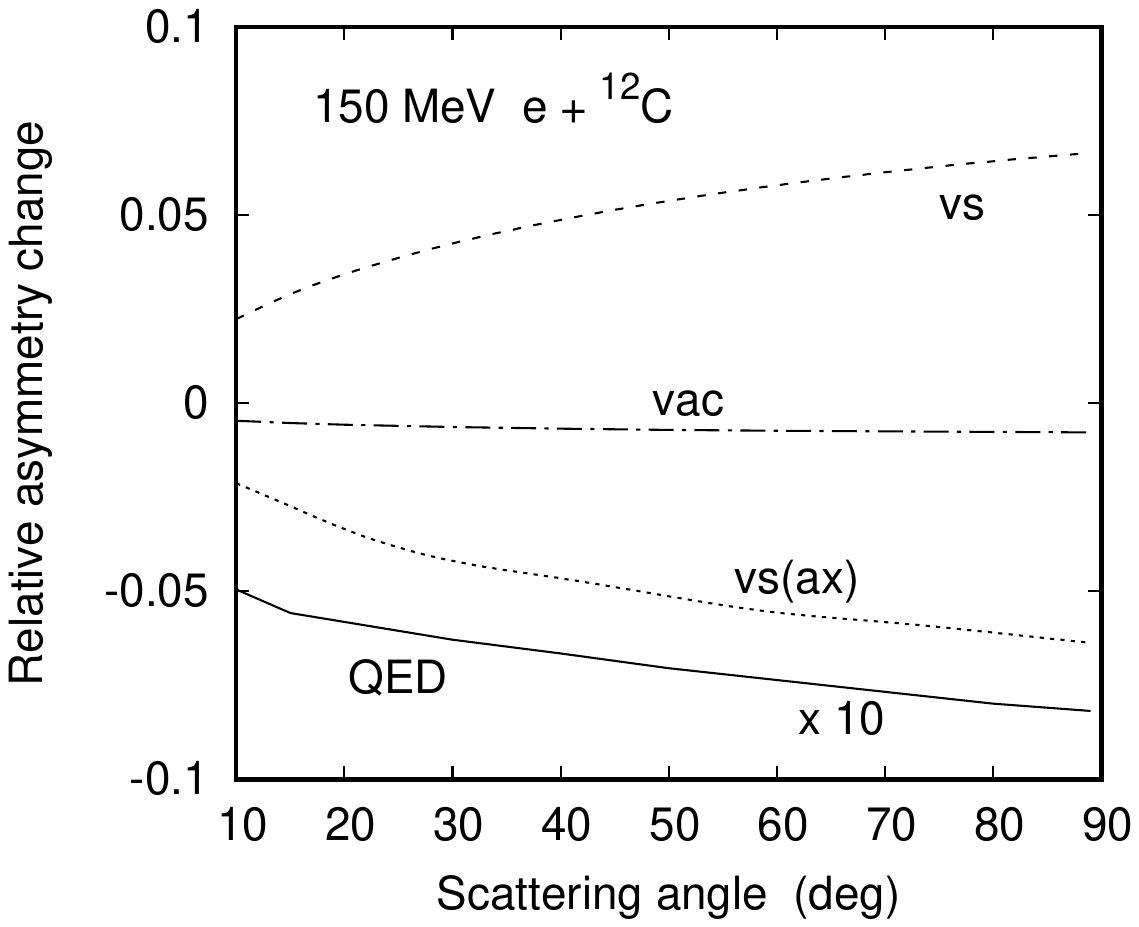}
\caption
{
Relative spin asymmetry change $dA_{\rm pv}$ in 150  MeV $e + ^{12}$C collisions as function of scattering angle $\vartheta_f$.
Shown are the QED changes by vacuum polarization ($dA_{\rm pv}^{\rm vac},\;-\cdot -\cdot -$), by the vector vs correction ($dA_{\rm pv}^{\rm vs},\;----$),
by the axial-vector vs correction $(dA_{\rm pv}^{\rm ax},\;\cdots\cdots$), and the total QED correction ($dA_{\rm pv}^{\rm QED}$, --------) multipied by a factor of 10.
}
\end{figure}

The angular distribution of the QED corrections to $A_{\rm pv}$ for 150 MeV electrons colliding with $^{12}$C, relevant for the planned low-energy MREx experiment at the MESA 
facility, is shown in Fig.4. 
It is seen that the counteraction of the two vs contributions holds in the whole forward hemisphere.
Given the scale of Fig.4, $dA_{\rm pv}^{\rm QED}$ basically coincides with $dA_{\rm pv}^{\rm vac}$,
a result which has also been found in \cite{RH26}.
Therefore $dA_{\rm pv}^{\rm QED}$ is enhanced by a factor of 10 in the figure in order to reveal its angular dependence, which is similar to the one of $dA_{\rm pv}^{\rm ax}$.

Results for the experimentally considered geometries of the four investigated targets, $^{12}$C, $^{27}$Al, $^{48}$Ca and $^{208}$Pb, are provided in Table \ref{tab1}.

\vspace*{0.5cm}
\begin{table}
 \caption{Changes $dA_{\rm pv}$ of the parity-violating asymmetry by the vector vs correction ($dA_{\rm pv}^{\rm vs}$), by the axial-vector vs correction ($dA_{\rm pv}^{\rm ax}$),  by the vacuum polarization ($dA_{\rm pv}^{\rm vac}$), by their sum, as well as the total change $dA_{\rm pv}^{\rm QED}$ from (\ref{7}). Also given is $A_{\rm pv}$ and its corrected value $A_{\rm pv}^{\rm QED}$.
The columns $2-7$ contain the results for $^{12}$C, $^{27}$Al, $^{48}$Ca and $^{208}$Pb, respectively, for the geometries indicated at the bottom of the Table.\label{tab1}}
 \centering
\begin{tabular}[t]{ccccccc}  
\hline\hline
 & $^{12}$C & $^{27}$Al  & $^{48}$Ca & $^{208}$Pb & $^{208}$Pb & $^{208}$Pb \\
\hline
$dA_{\rm pv}^{\rm vs}$ & $4.25 \times 10^{-2}$ & $5.98\times 10^{-2}$ & $6.29\times 10^{-2}$ & $4.82\times 10^{-2}$ & $4.78\times 10^{-2}$ & $5.08\times 10^{-2}$\\
$dA_{\rm pv}^{\rm ax}$ & $-4.19\times 10^{-2}$ & $-5.48\times 10^{-2}$ & $-5.52 \times 10^{-2}$& $-3.73 \times 10^{-2}$ & $-3.69\times 10^{-2}$ & $-3.74\times 10^{-2}$\\
$dA_{\rm pv}^{\rm vac}$ &$-6.30 \times 10^{-3}$ &  $-7.53\times 10^{-3}$ &$-7.95 \times 10^{-3}$ & $-7.98\times 10^{-3}$ & $-7.93\times 10^{-3}$& $-8.34\times 10^{-3}$\\
sum & $-5.70\times 10^{-3}$ & $-2.53\times 10^{-3}$ & $-2.5\times 10^{-4}$ &$2.92 \times 10^{-3}$ & $2.97 \times 10^{-3}$ & $5.06 \times 10^{-3}$\\
$d A_{\rm pv}^{\rm QED}$ &$-6.28\times 10^{-3}$ &$-6.45\times 10^{-3}$ && $1.04 \times 10^{-3}$ & $8.79\times 10^{-4}$ & $ 2.47 \times 10^{-3}$\\
\hline
$A_{\rm pv}$ & $5.131 \times 10^{-7}$ & $2.046 \times 10^{-6}$ & $2.303 \times 10^{-6}$& $6.364 \times 10^{-7} $ & $5.898 \times 10^{-7}$ & $7.178 \times 10^{-7}$\\
$A_{\rm pv}^{\rm QED}$ &$5.099\times 10^{-7}$&$2.033\times 10^{-6}$ & &$6.371 \times 10^{-7}$ & $5.903 \times 10^{-7}$ & $7.196 \times 10^{-7}$\\
\hline
$E$ & 150 MeV & 1157 MeV& 2180 MeV& 150 MeV & 953 MeV& 1063 MeV\\
$\vartheta_f$ & $30^\circ$ & $7.61^\circ$ & $4.51^\circ$ & $30^\circ$ & $4.7^\circ$ & $5^\circ$ \\
\hline\hline
\end{tabular}
\end{table}
\vspace{0.5cm} 

For energies in the GeV region and scattering angles around $5-7$ degrees, the change of $A_{\rm pv}$ by the vector vs perturbation is in absolute value somewhat larger than the axial-vector vs effect. In fact, for the $^{208}$Pb target, the combined effect is basically cancelled by the contribution from vacuum polarization, leading to an increase of the spin asymmetry.

For the light targets, the combined vs contribution is very small such that the total QED effect is mostly produced by vacuum polarization,
hence leading to a decrease of $A_{\rm pv}$.
In the computations, a Gaussian ground-state charge distribution \cite{deV} is used for $^{12}$C. 
Concerning the other elements, see \cite{RJ25}.
For $^{27}$Al, only the choice of the Fermi distribution (with parameters $r=3.07$ fm and $c=0.519$ fm)
leads to a reliable estimate of the QED corrections at energies as high as 1 GeV, while there is no convergence when using the Fourier-Bessel representation.
Due to mutual cancellations, the total QED results, $dA_{\rm pv}^{\rm QED}$, are  only accurate to about 10\%.
No convergence is obtained for $dA_{\rm pv}^{\rm QED}$ from the $^{48}$Ca target at 2.18 GeV with any choice of charge density.

For estimating all radiative corrections, also dispersion has to be considered.
For a moderate energy of 150 MeV our model, including explicitly the dominant low-lying nuclear excitations up to 30 MeV \cite{JR26}, is expected to give a reliable estimate. 
In order to judge the accuracy of our prediction for the $^{12}$C nucleus, we have included some more dipole excitations 
(most importantly, the isoscalar $1^-$ state at 33.64 MeV from the Skyrme SkM* nuclear model \cite{skm_star}), and we have found a 10\% reduction  
of the total dispersion correction at $30^\circ$
(from $8.53 \times 10^{-3}$ to $7.57 \times 10^{-3}$).
Note that for $^{12}$C dispersion counteracts $dA_{\rm pv}^{\rm QED}$, while for $^{208}$Pb it enhances the QED changes at this energy.
In conclusion, we have quantitatively and nonperturbatively calculated the QED corrections to the parity-violating asymmetry in polarized elastic electron scattering.
Our results for the individual QED contributions are in good agreement with those obtained in \cite{RH26}. 
However, deviations occur for the combined QED corrections which could be associated with the use of the Eikonal approximation in \cite{RH26}.
We confirm  that the axial-vector vs correction counteracts the contribution from the vector vs process.
Even more, the remaining correction 
 is  in case of $^{208}$Pb mostly compensated by vacuum polarization, such that the net effect of QED is very small. 
In particular, at forward angles and GeV electron beam energies the QED corrections are  below 1\%. Hence  they are non-relevant for the analysis of the PREx and CREx data, given their experimental uncertainties. 

For the lower energy of 150 MeV, the radiative corrections at $30^\circ$, apt to the planned MREx precision experiments, are
$1.86 \times 10^{-3}$ for $^{12}$C and $2.02 \times 10^{-3}$ for $^{208}$Pb, where also transient nuclear excitations are included (the dispersive corrections amout to $7.57 \times 10^{-3}$ for $^{12}$C and $8.98 \times 10^{-4}$ for $^{208}$Pb).
Hence these corrections have to be taken into account if the experimental error is below 0.3\%.

\ack{We wish to thank C.~J. Horowitz and B.~T. Reed for a fruitful correspondence.}

\funding{X.~R.~M. acknowledges support  by  MI-CIU/AEI/10.13039/501100011033  and  by FEDER  UE  through  grants  PID2023-147112NB-C22; and  through  the  “Unit  of  Excellence  Maria  de  Maeztu2025–2028”   award   to   the   Institute   of   Cosmos   Sci-ences,  grant CEX2024-001451-M. Additional support is provided  by  the  Generalitat  de  Catalunya  (AGAUR) through grant 2021SGR01095.}


\end{document}